\documentclass[conference]{IEEEtran}

\usepackage{theorem}
\usepackage{hyperref}
\usepackage{amssymb}
\usepackage{amsmath}

\usepackage{fancyvrb}
\fvset{fontsize=\footnotesize}
\DefineShortVerb{\|}
\DefineVerbatimEnvironment{v}{BVerbatim}{}

\newcommand\comment[1]{}

\newcommand\vide{\emptyset}
\newcommand\eg{{\em e.g.} }
\newcommand\ie{{\em i.e.} }

\newcommand\furl[1]{\footnote{\url{http://#1}}}

\renewcommand\prod{\mr{prod}}

\renewcommand\a{\rightarrow}
\newcommand\A{\Rightarrow}

\newcommand\ar{\a_\cR}

\newcommand\I[1]{[\![#1]\!]}

\newcommand\sle{\subseteq}

\newcommand\tlt{\lhd}

\newcommand\al{\alpha}
\renewcommand\b{\beta}

\renewcommand\l{\lambda}

\newcommand\s{\sigma}

\newcommand\vphi{\varphi}

\newcommand\mc[1]{{\mathcal{#1}}}

\newcommand\mr[1]{{\mathrm{#1}}}
\newcommand\mb[1]{{\mathbb{#1}}}

\newcommand\ms[1]{{\mathsf{#1}}}

\newcommand\bN{\mb{N}}

\newcommand\bZ{\mb{Z}}

\newcommand\cD{\mc{D}}

\newcommand\cF{\mc{F}}
\newcommand\cG{\mc{G}}

\newcommand\cN{\mc{N}}

\newcommand\cR{\mc{R}}

\newcommand\cT{\mc{T}}

\newcommand\cX{\mc{X}}

\renewcommand\sc{\ms{c}}

\renewcommand\sf{\ms{f}}
\newcommand\sg{\ms{g}}

\renewcommand\sl{\ms{l}}

\newcommand\sF{\ms{F}}

\newcommand\sT{\ms{T}}
\newcommand\sU{\ms{U}}

\newenvironment{rul}
  {$\begin{array}{rcl}}
  {\end{array}$}

\newenvironment{rew}[1][~~\a~~]
  {$\begin{array}{r@{#1}l}}
  {\end{array}$}
\newenvironment{rewc}[1][~~\a~~]
  {\begin{center}\begin{rew}[#1]}
  {\end{rew}\end{center}}

\newcounter{counter}

{\theorembodyfont{\rmfamily} 
  \newtheorem{dfn}[counter]{Definition}
  
  \newtheorem{thm}[counter]{Theorem}

}

\newenvironment{lstgeneric}[2]
  {\begin{list}{#1}{\topsep=.5mm\itemsep=.5mm\parsep=0mm%
    \itemindent=-3ex\labelsep=1ex\labelwidth=0ex #2}}
  {\end{list}}

\renewenvironment{c}{\begin{center}}{\end{center}}

\newcommand\WF{\ms{WF}}
\newcommand\DP{\ms{DP}}
\newcommand\Mon{\ms{Mon}}

\begin{document}

\title{Automated verification of termination certificates}

\author{
\IEEEauthorblockN{Fr\'ed\'eric Blanqui and Kim Quyen Ly}\\
\IEEEauthorblockA{INRIA, France\\
and\\
Institute of Software of the Chinese Academy of Sciences\\
4 South Fourth Street, Zhong Guan Cun\\
Beijing 100190, China}}

\maketitle

\begin{abstract}
In order to increase user confidence, many automated theorem provers
provide certificates that can be independently verified. In this
paper, we report on our progress in developing a standalone tool for
checking the correctness of certificates for the termination of term
rewrite systems, and formally proving its correctness in the proof
assistant Coq. To this end, we use the extraction mechanism of Coq and
the library on rewriting theory and termination called CoLoR.
\end{abstract}

\section{Introduction}
\label{sec-intro}


Being able to prove the correctness of a program is important,
especially for critical applications (banking, aeronautics, etc). But
this is generally undecidable. So, many different and complementary
approaches have been developed for tackling this problem: software
engineering methodologies, testing, model-checking, formal proof, etc.

Instead of trying to prove that every possible output of a program is
correct, one possible approach consists in making the tool provide, at
each run, an evidence that its output is correct. This certificate can
then be checked independently by another tool. Although it seems to
only move the problem from one program to the other, the certificate
verifier, there is in fact a gain in complexity. Typically, a program
which goal is to find a solution to some numerical or symbolic
problem, will use complex heuristics and optimizations, while checking
that the solution found is indeed correct is often much easier. For
instance, finding a boolean assignment satisfying some boolean formula
(SAT problem) is (in the worst case) exponential in the number of
boolean variables, while verifying the correctness of a given
assignment (the certificate) is linear in the size of the formula.


Since certificate verifiers are simpler programs, they are more easily
amenable to a complete formalization and proof using some proof
assistant tool. In fact, various such tools (\eg Coq \cite{coq}) are
themselves based on this two-level approach: they are composed of a
small and hopefully safe kernel responsible of checking the
correctness of proofs, and a proof development environment providing
unsafe proof tactics and decision procedures for building step by step
proofs that, in the end, have to be checked by the kernel to be
included in the proof database.

\smallskip

Termination, that is, the fact that a program eventually provides an
output to the user, is an important property that is also undecidable
\cite{dauchet88mfcs}. Term rewriting
\cite{dershowitz90book,terese03book} is a simple yet very general
programming paradigm and framework, based on the notion of rewrite
rule, that generalizes or in which to easily encode other programming
paradigms like functional or logic programs. Examples of programming
languages based on rewriting are \cite{cafeobj,elan,tom,maude}. A few
years ago, a formal language called CPF \cite{cpf} has been developed
that defines a notion of certificate for the termination of term
rewrite systems.


\smallskip

In this paper, we consider the problem of developing a standalone tool
for checking the correctness of CPF certificates, and formally proving
its correctness. In \cite{blanqui11mscs}, the first author describes a
CPF verifier called
Rainbow\footnote{\url{http://color.inria.fr/rainbow.html}} based on
the following architecture: a compiler (written in OCaml \cite{ocaml})
from CPF to Gallina, the language of the Coq proof assistant
\cite{coq}, generates a Gallina script that is then checked by Coq
itself using the Coq library CoLoR \cite{blanqui11mscs}. This
architecture has some advantages: it provides a way to automatically
generate Coq representations of term rewrite systems and termination
arguments that can be used for proving the termination of Coq
functions. Indeed, in Coq, no function can be defined without proving
its termination, because allowing non-terminating functions would make
proof verification undecidable. But this architecture has also some
disadvantages. First, compared to more standard programming languages,
computation in Coq is very slow (and indeed too slow to check some
complex termination certificates). Second, the compiler from CPF to
Coq is not proved and can thus introduce errors not present in the
certificate.


Here, we consider a different architecture based on Coq's ability to
generate OCaml \cite{ocaml}, Haskell \cite{haskell} or Scheme
\cite{scheme} programs equivalent to the functions defined in it
\cite{letouzey04phd}. It consists in defining the CPF verification
program directly in Coq (except the parsing part), and prove its
correctness. Then, Coq's extraction mechanism provides us with an
OCaml, Haskell or Scheme standalone program that can be compiled and
efficiently executed independently of Coq or the CoLoR library.

A similar approach has been undertaken successfully for the CPF
verifier CeTA \cite{ceta} with the proof assistant Isabelle/HOL
\cite{nipkow02lncs,haftmann09phd}, which implements classical
higher-order logic with the axiom of choice \cite{church40jsl}. Here,
we want to test this approach in the proof assistant Coq, which
implements an extension of intuitionist higher-order logic
\cite{coquand88ic,paulin93tlca}, and by using the CoLoR library.


The first problem to address is the representation in Coq of CPF
certificates. The second one is the formalization and proof of the CPF
verifier program using the Coq library on rewriting theory and
termination called CoLoR \cite{blanqui11mscs}. In particular, it
requires to translate the CPF data structures into the data structures
used in CoLoR.


\smallskip

This paper is organized as follows. In section \ref{sec-trs}, we
introduce term rewriting systems and give some examples of termination
techniques used in current automated termination provers. In section
\ref{sec-cpf}, we describe the formal language CPF for termination
certificates used in the international competition of automated
termination provers \cite{tc}. In section \ref{sec-coq}, we introduce
the proof assistant Coq and how to formalize and prove the correctness
of a certificate verifier in it. In section \ref{sec-cpf-in-coq}, we
give some details on the representation of certificates in
Coq. Finally, in section \ref{sec-prf}, we give some details on the
formalization and proof of the verifier using the CoLoR library.

\section{Term rewrite systems and their termination}
\label{sec-trs}


We first recall what is rewriting: ``rewrite systems are directed
equations used to compute by repeatedly replacing subterms of a given
formula with equal terms until the simplest form possible is
obtained'' \cite{dershowitz90book}. More formally:

\begin{dfn}[Term rewrite system]
Let $\cX$ be an infinite set of {\em variables}. Given a set $\cF$ of
{\em function symbols} (disjoint from $\cX$) and an arity function
$\al:\cF\a\bN$, the set $\cT(\cF,\cX)$ of (first-order) {\em terms}
over $\cF$ and $\cX$ is the smallest set containing $\cX$ and such
that, if $\sf\in\cF$ and $t_1,\ldots,t_{\al(\sf)}$ are terms, then
$\sf(t_1,\ldots,t_{\al(f)})$ is a term.

A {\em substitution} $\s$ is a map from variables to terms that is
extended to terms in the obvious way ($x\s=\s(x)$ and
$\sf(t_1,\ldots,t_n)\s=\sf(t_1\s,\ldots,t_n\s)$). A {\em context} $C$
is a term with a unique occurrence of a distinguished variable $[]$,
which substitution by $u$ is written $C[u]$. A (rewrite) rule is a
pair of terms written $l\a r$. The {\em rewrite relation} $\ar$
generated by a set $\cR$ of rules is the smallest relation containing
$\cR$ and stable by substitution ($t\ar u\A t\s\ar u\s$) and context
($t\ar u\A C[t]\ar C[u]$).

A relation $\a$ terminates (or is well-founded, or noetherian) if
there is no infinite sequence $t_0\a t_1\a\ldots$
\end{dfn}

A simple example of rewrite system is given by the addition on unary
natural numbers:
\[\ms{add}(\ms{zero},x)\a x\quad
\ms{add}(\ms{succ}(x),y)\a \ms{succ}(\ms{add}(x,y))\]


The termination of a TRS is undecidable in general, even with a single
rule \cite{dauchet88mfcs}. So, there has been active research for
finding powerful sufficient conditions. An important one consists in
interpreting function symbols by monotone polynomials on natural
numbers $\bN$ \cite{lankford79tr,contejean05jar}:

\begin{thm}[Polynomial interpretation]
Let $\cR$ be a TRS and $\vphi$ be a function mapping a polynomial
$\vphi_\sf\in\bZ[X_1,\ldots,X_n]$ to each function symbol $\sf$ of
arity $n$. Given a valuation $\al:\cX\a\bN$, let
$\I{x}^\vphi_\al=\al(x)$ and $\I{\sf(t_1,\ldots,t_n)}^\vphi_\al=
\vphi(\sf)(\I{t_1}^\vphi_\al,\ldots,\I{t_n}^\vphi_\al)$ be the
interpretation of terms in $\bZ$ induced by $\vphi$, and $t>_\vphi u$
if, for all $\al$, $\I{t}^\vphi_\al>_\bN\I{u}^\vphi_\al$, the
well-founded ordering on terms induced by $\vphi$.

If every $\vphi_\sf$ is monotone in every $x_i$,
${\cR_1}\sle{>_\vphi}$ and $\a_{\cR_2}$ terminates, then
$\a_{\cR_1\cup\cR_2}$ terminates.
\end{thm}

For instance, the previous system can be proved terminating by using
the following polynomial interpretation on $\bN$:
\[\vphi_\ms{add}(x,y)=2x+y\quad \vphi_\ms{succ}(x)=x+1\quad \vphi_\ms{zero}=1\]

Indeed, for the first rule, we have $2(1)+x>_\bN x$ and, for the
second rule, we have $2(x+1)+y>_\bN (2x+y)+1$, whatever are the values
of $x,y\in\bN$.

Another very important method, at the basis of all current TRS
termination provers, consists in transforming a TRS into a dependency
pair (DP) problem \cite{arts00tcs}:

\begin{dfn}[Dependency pair]
Given a set of symbols $\cF$, the set
$\cF^\sharp=\cF\uplus\{\sf^\sharp\mid\sf\in\cF\}$ which consists of
the disjoint union of $\cF$ with some copy of $\cF$, is the set of
marked and unmarked symbols ($\sf^\sharp$ is taken to be of same arity
as $\sf$). Given a set $\cR$ of rules, a symbol $\sf$ is said {\em
  defined} if there is a rule whose left hand-side is of the form
$\sf(l_1,\ldots,l_n)$. Let $\cD(\cR)$ be the set of defined
symbols. The set of {\em dependency pairs} $\DP(\cR)$ is then the set
of marked rules
$\sf^\sharp(l_1,\ldots,l_n)\a\sg^\sharp(r_1,\ldots,r_p)$ such that
$\sf(l_1,\ldots,l_n)\a r\in\cR$ for some $r$, $\sg(r_1,\ldots,r_p)$ is
a subterm of $r$ not occurring in some $l_i$, and $\sg$ is
defined. The {\em dependency graph} whose nodes are $\DP(\cR)$ has an
edge between $(l_1,r_1)$ and $(l_2,r_2)$ if there are two
substitutions $\s_1$ and $\s_2$ such that $r_1\s_1\ar^*l_2\s_2$.
\end{dfn}

Indeed, $\ar$ terminates on $\cT(\cF,\cX)$ iff the composition of the
reflexive-transitive closure of $\ar$ with the closure by substitution
of $\DP(\cR)$, written $\ar^*\a_{\DP(\cR)h}$, terminates on
$\cT(\cF^\sharp,\cX)$. Intuitively, dependency pairs generalizes the
notion of recursive calls and call graph in functional programming
\cite{thiemann05aaecc}. Interpretations in a well-founded domain are
easily extended to deal with this more general kind of
relations. Moreover, since we only consider the closure by
substitution of $\DP(\cR)$, only one dependency pair need to strictly
decrease in every cycle or, more simply, in every connected component
of the dependency graph. This allows to split a DP problem into
various independent DP sub-problems \cite{hirokawa07ic}.

For instance, in our simple example, there is only one dependency
pair, $\ms{add}^\sharp(\ms{succ}(x),y)\a\ms{add}^\sharp(x,y)$, the
termination of which can be proved by taking
$\vphi_{\ms{add}^\sharp}(x,y)=x$.\\

\section{Termination certificates}
\label{sec-cpf}


The theorem on polynomial interpretation can be described as a
conditional deduction rule on termination problems:

\begin{c}
(rule-removal-PI)
$\cfrac{\Mon(\vphi)\quad {\cR_1}\sle{>_\vphi}\quad \WF(\a_{\cR_2})}
{\WF(\a_{\cR_1\cup\cR_2})}$
\end{c}

\noindent
where $\Mon(\vphi)$ means that every $\vphi_\sf$ is monotone in every
$x_i$, ${\cR_1}\sle{>_\vphi}$ that every rule of $\cR_1$ is strictly
decreasing in the interpretation, and $\WF(\a_{\cR_2})$ that
$\a_{\cR_2}$ terminates (is well-founded).

Similar conditional deduction rules can be written for most if not all
termination methods used in current termination provers
\cite{giesl04lpar}. Hence, a termination proof can be described by a
deduction tree obtained by composing deduction rules like
(rule-removal-PI) and axioms like:

\begin{c}
(empty) $\cfrac{\cR=\vide}{\WF(\ar)}$
\end{c}


For the international competition of automated termination provers
\cite{tc}, a formal language called CPF \cite{cpf} has been
collectively defined for representing such deduction trees. It is
given as an XML Schema or XSD file \cite{xsd1,simeon03popl}. An XSD
file is like a grammar: it describes the set of XML files that are
admissible. XML is a well established W3C text file standard
\cite{xml} for describing tree-structured data. For instance, in CPF,
a rewrite rule has to be described by the following XML text:

\begin{c}
\begin{v}
<rule><lhs>...</lhs><rhs>...</rhs></rule>
\end{v}
\end{c}

It represents a labeled tree, which root is labeled by the tag |rule|,
having two sub-trees: the first one describes the rule left hand-side
and has its root labeled by the tag |lhs|, and the second one
describes the rule right hand-side and its root is labeled by the tag
|rhs|. The XML Schema language (which is a subset of XML) allows to
describe some set of valid XML texts by declaring what are the
possible labeled trees. For instance, the XSD type used in CPF for
rewrite rules is:

\begin{c}
\begin{v}
<xs:element name="rule">
  <xs:complexType>
    <xs:sequence>
      <xs:element name="lhs">
        <xs:complexType>
          <xs:group ref="term"/>
        </xs:complexType>
      </xs:element>
      <xs:element name="rhs">
        <xs:complexType>
          <xs:group ref="term"/>
        </xs:complexType>
      </xs:element>
    </xs:sequence>
  </xs:complexType>
</xs:element>
\end{v}
\end{c}

The main type constructors allowed in XSD are, informally:

\begin{itemize}
\item|element|: if $T$ is an XSD type and $x$ is a string, then
  |<element name="|$x$|">|$T$|</element>| denotes the set of trees
  which root is labeled by $x$ and which children belong to the set of
  trees corresponding to $T$.

\item|sequence|: if $T_1,\ldots,T_n$ are XSD types, then
  |<sequence>|$T_1\ldots T_n$|</sequence>| denotes\footnote{In the
    complete definition, every type $T_i$ can be equipped with two
    attributes $a\in\bN$ and $b\in\bN\cup\{\infty\}$ specifying the
    minimum and maximum numbers ($\infty$ meaning arbitrary) of
    children of type $T_i$.} the set of tuples of trees
  $(t_1,\ldots,t_n)$ such that $t_1$ is of type $T_1$, \ldots, $t_n$
  is of type $T_n$.

\item|choice|: if $T_1,\ldots,T_n$ are XSD types, then
  |<choice>|$T_1\ldots T_n$|</choice>| denotes the union of the sets
  of trees corresponding to $T_1,\ldots,T_n$.
\end{itemize}

\section{Formalization and proof of a certificate verifier in Coq}
\label{sec-coq}


The Coq proof assistant \cite{coq} is a tool that allows one to
formally define mathematical objects and prove statements about
them. It has been successfully used in the certification of various
important applications, either industrial: a JavaCard platform
\cite{barthe02amast} or a C compiler \cite{leroy09cacm}, or
academical: the four color theorem \cite{gonthier07ascm} or Kepler's
conjecture \cite{hales09dcg}.

It is based on an extension of Girard' system F \cite{girard88book}
and Martin-L\"of type theory \cite{martinlof84book}, called the
calculus of inductive constructions \cite{coquand88ic,paulin93tlca}.
It allows function definitions by pattern-matching \cite{cornes97phd}
and provides a programmable proof tactic language
\cite{delahaye00lpar}, various decision procedures, and other
important features like modules, type classes, etc.

It is therefore possible to define in Coq an inductive data type |cpf|
for representing CPF predicates, a boolean function
|check:trs->cpf->bool| verifying the correctness of a certificate wrt
a termination problem, and formally prove that this function is
correct, that is, in Coq syntax:

\begin{c}
\begin{v}
Theorem check_is_correct:
  forall R x, check R x = true -> WF (red R).

Proof. ... Qed.
\end{v}
\end{c}

In fact, in order to provide useful error messages if a certificate
appears to be incorrect, to deal with certificates that the verifier
does not know how to handle yet (there many different certificates in
CPF and it is a really huge work to handle all of them), instead of a
boolean output, we use an error monad \cite{wadler92mscs}. And since
many auxiliary functions are necessary for translating CPF data
structures into CoLoR data structures, we use a polymorphic error
monad:

\begin{c}
\begin{v}
Inductive result (A : Type) : Type :=
| Ok : A -> result A
| Ko : error -> result A.

Definition term : cpf_term -> result color_term :=
  ...

Theorem check_is_correct:
  forall R x, check R x = Ok unit -> WF (red R).
\end{v}
\end{c}

Finally, since Coq includes a typed $\l$-calculus with inductive data
types and pattern-matching, the extraction of ML-like function
definitions \cite{harper86tr} from Coq to OCaml \cite{letouzey04phd}
is almost straightforward\footnote{Note however that parallel
  pattern-matching and pattern-matching with patterns of depth greater
  than 1 are not primitive in Coq. They are compiled into sequences of
  non-parallel pattern-matching with patterns of depth 1, leading to
  important code duplication in some cases.}\footnote{This is however
  not the case of more complex Coq constructions
  \cite{letouzey04phd,glondu12phd}.}  and looks about the same since
Coq syntax is very close to OCaml syntax.

\section{Parsing and Coq representation of certificates}
\label{sec-cpf-in-coq}


The CPF format is extended every year with new certificates and can be
modified sometimes. In Rainbow, the data type used for representing
certificates internally and the parsing function used to create a
value of this data type from a text file are written by hand (the
parsing function uses the XML-Light library \cite{xml-light}). This is
a possible source of errors and is time-consuming.

To avoid these problems, we developed a compiler from XSD to Coq and
OCaml that, from an XSD file, generates a Coq file (and hence an OCaml
file after extraction from Coq) providing a data type definition for
representing XML data valid wrt the given XSD file, and an OCaml file
providing a parsing function for this data type (also based on
XML-Light). This compiler is not intended to cover all aspects of XSD
but only the one used in CPF.

The XSD type constructors described above are translated to standard
OCaml data structures as follows (with some optimizations):

\begin{itemize}
\item|sequence|: tuple or list (an optional child being mapped to the
  OCaml |option| type);
\item|choice|: data type with a constructor for each case.
\end{itemize}

For instance, in CPF, the type for function symbols is defined as
follows:

\begin{c}
\begin{v}
<xs:group name="symbol">
  <xs:choice>
    <xs:element ref="name"/>
    <xs:element name="sharp">
      <xs:complexType>
        <xs:sequence>
          <xs:group ref="symbol"/>
        </xs:sequence>
      </xs:complexType>
    </xs:element>
    <xs:element name="labeledSymbol">
      <xs:complexType>
        <xs:sequence>
          <xs:group ref="symbol"/>
          <xs:group ref="label"/>
        </xs:sequence>
      </xs:complexType>
    </xs:element>
  </xs:choice>
\end{v}
\end{c}

\noindent
where |<group name="|$x$|">| is a way in XSD to introduce a type
definition that can be referred to by $x$. This XSD type is translated
by our compiler to the following inductive OCaml data type:

\begin{c}
\begin{v}
type symbol =
| Symbol_name of name
| Symbol_sharp of symbol
| Symbol_labeledSymbol of symbol * label
\end{v}
\end{c}

Other solutions could be chosen. Note however that not every OCaml
value corresponds to an XML file validating CPF. To do so, we would
need to use private data types \cite{blanqui07esop} or a stronger type
system like the one of CDuce \cite{cduce,frisch08jacm}.


More importantly, in XSD, type definitions are unordered and a type
definition can refer to types defined later in the file. This is not a
problem in itself for OCaml or Coq since these languages support
mutually defined types too. However, if CPF is represented in Coq as a
single big set of mutually defined types, then Coq will generate a
single big induction principle for all types that will be very
difficult to use in proofs. It is therefore better to have as many
minimal sets of mutually defined types as possible. And because in Coq
and OCaml, the type names used in a type definition can only refer to
type names of the same set of mutually defined types or to previously
defined types, it is necessary to order the XSD type definitions wrt
their dependencies:

\begin{dfn}[Type dependency relation]
For our purpose\footnote{This is the class of OCaml types to which XSD
  types are compiled.}, we can consider that a type $\sT$ is defined
by a finite set of constructors the arguments of which are of type
$\sT_1,\ldots,\sT_n$ respectively. Then, we say that a type $\sT$
depends on a type $\sU$, written $\sU\tlt\sT$, if there is a
constructor of $\sT$ having an argument of type $\sU$. And we say that
a type $\sU$ must be defined before a type $\sT$, written
$\sU\preceq\sT$, if $(\sU,\sT)$ is in the reflexive and transitive
closure of $\tlt$. We then denote by $\simeq$ the symmetric closure of
$\preceq$ (it is an equivalence relation), and by
${\prec}={\preceq-\simeq}$ its strict part.
\end{dfn}

The minimal sets of mutually dependent types correspond then to the
equivalence classes of the $\simeq$ equivalence relation, and these
classes can be ordered topologically by using $\prec$.

\section{Definition and proof of a termination certificate verifier in Coq}
\label{sec-prf}


The first problem to address is the translation of CPF data structures
for symbols, terms, rules, polynomials, etc. to the corresponding
CoLoR data structures. In fact, this is more or less straightforward
except for terms.

In CoLoR, every definition or theorem is parametrized by a given
signature:

\begin{c}
\begin{v}
Record Signature : Type := mkSignature {
  symbol :> Type;
  arity : symbol -> nat;
  beq_symb : symbol -> symbol -> bool;
  beq_symb_ok :
    forall x y, beq_symb x y = true <-> x = y }.
\end{v}
\end{c}

\noindent
providing the set of symbols, their arity and a boolean function on
symbols ensuring that equality on symbols is decidable.

Then, new sets are introduced when needed, like it is the case for
marked symbols in the dependency pairs transformation. Moreover, some
termination techniques may change the arity of symbols. For instance,
arguments filtering \cite{arts00tcs} may transform a TRS where $\sf$
is of arity $n\ge 1$ into a TRS where $\sf$ is of arity $n-1$ by
removing the first argument of $\sf$ in every rule where $\sf$
occurs.

Hence, in CoLoR, the set of symbols and their arity may evolve
dynamically during the verification of a certificate, and differently
wrt the deduction branch followed (a certificate has a tree
structure), while, in CPF, there is only one big type for all the
possible symbols. Defining a function for converting a CPF term into a
CoLoR term following the same dynamic would be complicated.

Instead, we use the fact that the CPF type for symbols include all
possible symbols that can be generated in the course of a
verification, and chose the CPF type itself for the set of CoLoR
symbols. Hence, only the arity function needs to evolve dynamically.
Note that this is correct to do so since signature extension reflects
termination: given a set $\cR$ of rules on $\cT(\cF,\cX)$, if
$\cF\sle\cG$, then $\ar^\cF$ terminates iff $\ar^\cG$ terminates,
where $\ar^A$ is the relation generated by $\cR$ on $\cT(A,\cX)$
\cite{ohlebusch93eatcs}.

As a consequence, we need to translate CoLoR data structures for new
symbols back into the |cpf| data type. To prove that this
transformation reflects termination, we use the following theorem on
signature morphisms formalized in CoLoR:

\begin{thm}[Signature morphism]
Let $\cF$ and $\cG$ be two sets of symbols whose arity functions are
$\al$ and $\b$ respectively, and let $\vphi$ be a map from $\cF$ to
$\cG$ that respects arities, \ie forall $\sf\in\sF$,
$\b_{\vphi(\sf)}=\al_\sf$. The map $\vphi$ then naturally extends to
terms as follows: $\vphi(x)=x$ and
$\vphi(\sf(t_1,\ldots,t_n))=\vphi(\sf)(\vphi(t_1),\ldots,\vphi(t_n))$.

If $\cR$ is a set of rules on $\cT(\cF,\cX)$ and $\a_{\vphi(\cR)}$
terminates on $\cT(\cG,\cX)$, then $\ar$ terminates on $\cT(\cF,\cX)$.
\end{thm}

Note that no property is required for $\vphi$ other than to respect
arities. In particular, it does not need to be injective.

We now show how this applies on the DP transformation. Let $\cF$ be
the set of symbols corresponding to the data type |symbol| defined in
the previous section. To simplify, we do not consider the constructor
|Symbol_labeledSymbol|. So, $\cF$ can be seen as the solution of the
equation $X=\cN\uplus\{\sf^@\mid\sf\in X\}$, where $\cN$ is the set of
values of type |name| and $@$ stands for the constructor
|Symbol_sharp| to distinguish it from the symbol $\sharp$ used in the
DP transformation. Let $\cR$ be a set of rules on $\cF$ with no symbol
of the form $\sf^@$ such that $\ar^*\a_{\cD h}$ terminates, where
$\cD=\vphi(\DP(\cR))$ with $\vphi(\sf^\sharp)=\sf^@$ and
$\vphi(\sf)=\sf$ otherwise. Then, by the theorem on signature
morphisms, $\ar^*\a_{\DP(\cR)h}$ terminates and, by the DP theorem,
$\ar$ terminates.

\section{Conclusion}
\label{sec-conclu}

We started to develop a standalone tool for verifying the correctness
of termination certificates for term rewrite systems
\cite{dershowitz90book} following the CPF format \cite{cpf} used in
the international competition of automated termination provers
\cite{tc}, and formally prove its correctness in the proof assistant
Coq \cite{coq} using the Coq library on rewriting theory and
termination CoLoR \cite{blanqui11mscs} and Coq extraction mechanism
\cite{letouzey04phd}.


We first developed a simple compiler for generating a Coq data type
definition for representing XML Schema data types, and an XML parser
for CPF. We also defined and proved in Coq a small verifier for two
important termination techniques: dependency pairs \cite{arts00tcs}
and polynomial interpretations \cite{contejean05jar}. But much more
has to be done to be able to compete with the verifier CeTA developed
in the proof assistant Isabelle/HOL \cite{ceta}.

\bibliographystyle{IEEEtran}
\bibliography{main}

\end{document}